\documentclass[preprint,aps,showpacs,eqsecnum,nofootinbib]{revtex4}

\usepackage{epsfig}
\usepackage{amsmath}
\usepackage{amsfonts}
\usepackage{amssymb}
\usepackage{mathrsfs}
\usepackage{color}

\def\bc{\begin{center}}
\def\nno{\nonumber}
\def\ec{\end{center}}
\def\be{\begin{eqnarray}}
\def\ee{\end{eqnarray}}

\newcommand{\omits}[1]{}

\newcommand{\At}[1]{\bigg|_{#1}}
\newcommand{\vect}[1]{\mbox{\boldmath{$ #1$}}}
\newcommand{\const}{\mathrm{const}}

\definecolor{dyellow}{rgb}{1.,0.8,.0}
\definecolor{myblue}{rgb}{.1,.1,.7}
\definecolor{dcyan}{rgb}{.0,.6,.6}
\definecolor{dmagenta}{rgb}{0.6,0.0,0.6}
\definecolor{brown}{rgb}{0.6,0.2,0.}
\definecolor{darkblue}{rgb}{.0,.0,0.5}
\definecolor{darkred}{rgb}{0.75,0.0,0.0}
\definecolor{orange}{rgb}{1.,.6,.0}
\definecolor{dorange}{rgb}{0.8,.4,.0}
\definecolor{darkgreen}{rgb}{0.0,0.6,0.0}
\definecolor{purple}{rgb}{.4,.0,.4}

\def\blue{\color{blue}}
\def\yellow{\color{yellow}}

\def\darkgreen{\color{darkgreen}}

\def\La{\Lambda}
\def\Si{\Sigma}
\def\Om{\Omega}

\def\ga{\gamma}
\def\dl{\delta}
\def\eps{\epsilon}

\def\la{\lambda}

\def\si{\sigma}

\def\d#1#2{\frac{\displaystyle #1}{\displaystyle #2}}
\def\r{\partial}

\newcommand\btd{\raise 2pt
\hbox{$\hat\bigtriangledown$}\hskip 1.5pt}
\newcommand\bt{\raise 2pt
\hbox{$\bigtriangledown$}\hskip 1.5pt}

\begin{document}


\title{Newton-Hooke limit of Beltrami-de Sitter spacetime,\\
 principle of Galilei-Hooke's relativity
 and \\ postulate on Newton-Hooke universal time}

\author{{Chao-Guang Huang}$^{1}$}
\email{huangcg@mail.ihep.ac.cn}
\author{{Han-Ying Guo}$^{2}$} \email{hyguo@itp.ac.cn}
\author{{Yu Tian}$^{3,2}$}
\email{ytian@bit.edu.cn}
\author{{Zhan Xu}$^{4}$}
\email{zx-dmp@mail.tsinghua.edu.cn}
\author{{Bin Zhou}$^{5,2}$} \email{zhoub@itp.ac.cn}

\affiliation{%
${}^1$ Institute of High Energy Physics, Chinese Academy of
Sciences, P.O. Box 918-4, Beijing
   100049, China,}

\affiliation{%
${}^2$ Institute of Theoretical Physics,
 Chinese Academy of Sciences,
 P.O.Box 2735, Beijing 100080, China,}

\affiliation{%
${}^3$ Department of Physics,
 Beijing Institute of Technology,
 Beijing 100081, China,}

\affiliation{%
${}^4$ Physics Department, Tsinghua University, Beijing
   100084, China,}

\affiliation{%
${}^5$ Department of Physics, Beijing Normal University, Beijing
   100875, China,}


\vskip 15mm

\begin{abstract}
Based on the Beltrami-de Sitter spacetime, we present the
Newton-Hooke model under the Newton-Hooke contraction of the $BdS$
spacetime with respect to the transformation group, algebra and
geometry. It is shown that in Newton-Hooke space-time, there are
inertial-type coordinate systems and inertial-type observers,
which move along straight lines with uniform velocity. And they
are invariant under the Newton-Hooke group. In order to determine
uniquely the Newton-Hooke limit, we propose the Galilei-Hooke's
relativity principle as well as the postulate on
Newton-Hooke universal time. 
All results are readily
extended to the Newton-Hooke model as a contraction of
Beltrami-anti-de Sitter spacetime with negative cosmological
constant. 
\end{abstract}

\bigskip


\maketitle

\tableofcontents

\section{Introduction}

It is well known that there are different kinds of kinematics on
homogeneous (3+1)-d space-times \cite{Bacry}, and all of them can
be contracted from Minkowski, de Sitter or anti-de~Sitter
spacetimes under group contraction \cite{IW},
respectively\footnote{In the present paper, we use {\it spacetime} or
{\it space-time} to denote a unified manifold of space and time with or without
an invariant metric, respectively.}. %
Among them, Newton-Hooke (NH) space-times ${\cal NH}_\pm$ introduced in \cite{Bacry} are of special
interest as a kind of non-relativistic cosmological models
in \cite{ABCP, Gao, Gibbons}.

In this paper, we present a model for the NH space-time
${\cal NH}_+$ based on the Beltrami-de~Sitter ($BdS$) spacetime
\cite{BdS, Lu, LZG}, denoted by ${\cal B}_\Lambda$. We find that in
${\cal NH}_+$ there exist a kind of special coordinate systems
in which the test particles, which none force acts upon, move at constant
coordinate velocities along straight lines. In the coordinates, these
particles look like free particles rather than the particles driven by a
repulsive force provided by the positive cosmological constant in other
coordinates, say, the ones contracted from the {\it static} $dS$
universe \cite{Gibbons}. At the first glance, this property seems
very strange. Why in ${\cal NH}_+$ contracted from the $dS$
spacetime with constant curvature there exist such a kind of
uniform-velocity motions? The basic reason is, in fact, that there
exist a kind of inertial-type motions in the Beltrami model
of the $dS$ spacetime. Such an inertial-type property comes in
${\cal NH}_+$  from the $BdS$ spacetime ${\cal B}_\Lambda$.

Among various models of $dS$ spacetimes, the Beltrami model is an
important one in which $dS$ spacetime is in analog with Minkowski
spacetime in the sense that there are inertial-type coordinates,
inertial-type motions and they are invariant under fractional
linear transformations of a 10-parameter group --- $dS$~group
$SO(1,4)$. It is precisely the Beltrami model \cite{beltrami}
of a 4-hyperboloid ${\cal S}_\Lambda$ in the 5-d Minkowski
spacetime. In  ${\cal B}_\Lambda$, a set of Beltrami coordinate
systems covers ${\cal S}_\Lambda$ patch by patch and test
particles and light signals move along the timelike and null
geodesics, respectively, at {\it constant} coordinate velocities.
Therefore, they look like in free motions in a spacetime without
gravity. Thus, the Beltrami coordinates and observers ${\cal
O}_{B}$  in these systems may be regarded as of ``global
inertial-type".   This is why there is such a merit of the
NH contraction from $BdS$ spacetime that in ${\cal
NH}_+$ there remain inertial-type coordinates and inertial-type
observers.

This also shows that the usual algebraic definition of the NH
limit cannot determine the contraction of geometry uniquely
because of free choice of coordinates if coordinates have no
fundamental physical meaning. In particular, for the same kind of
coordinates the contraction depends on what dimension of the
coordinates might be chosen. In order to determine the unique NH
limit, we should supplement certain physical conditions. We argue
that the suitable physical conditions should be the counterpart of
Galilean relativity principle in ${\cal NH}_+$, which is named the
Galilei-Hooke's relativity principle, and the postulate on
Newton-Hooke universal time $t$, which satisfies $|t|{ < }
\nu^{-1}=R/c$ and the latter is invariant under the Newton-Hooke
contraction  of $c, R \to \infty$, i.e. the limit of infinite
signal-velocity $c$ and the curvature radius $R$ of the $BdS$
spacetime. The Newton-Hooke universal time is similar to the
Newton's universal time in Newtonian mechanics in the sense that
it is separated { from} space in the metric.

We also find that there is something interesting in ${\cal NH}_+$,
which is closely related to the violation of Euclid's fifth axiom.
In Newtonian mechanics and special relativity, a particle in
inertial motion can be stationary in only one inertial frame. If
an inertial frame $S'$ has a relative velocity in another one,
$S$, the relative velocity will be constant at every time and
every point. In ${\cal NH}_+$, however, it is possible that a
particle in inertial-type motion can be at rest in two different
inertial-type frames. In this case the velocity of $S'$ relative
to $S$ will be different from point to point.

This paper is organized as follows. In Sec.~\ref{BdS}, we briefly
review $BdS$ spacetime. In Sec.~\ref{NHlimit-IM} we study the
contraction of transformation group and geometry of $BdS$
spacetime in the NH limit. We also study motion of particles and
light signals in the NH limit of $BdS$ spacetime. In Sec.
\ref{unique-RP}, we show why the usual limiting procedure cannot lead to the unique NH limit. We propose the Galilei-Hooke's
relativity principle and the  postulate on
Newton-Hooke universal time  to uniquely determine the contraction
of geometry in the NH limit.  In Sec.~\ref{sect:GRef}, we discuss the
behaviors of inertial frames in ${\cal NH}_+$ that are different
from Newtonian mechanics and special relativity. Geometric
diagrams are also given in this section. Finally, we end with a
few concluding remarks. In appendix~\ref{appdx:conncurvt}, we list
some results of the connection and curvature of $BdS$ spacetime
and their contraction in the NH limit. In
appendix~\ref{appdx:limit} we use the 5-d Minkowski spacetime to
show why the NH limit is not unique.

\section{The Beltrami-de Sitter Spacetime and Inertial-type Motion\label{BdS}}

\subsection{The Beltrami-$dS$ Spacetime\label{sect:BdS}}

We start with a 4-d hyperboloid ${\cal S}_\Lambda$ embedded in a
5-d Minkowski spacetime with $\eta_{AB}= {\rm diag}(1, -1, -1, -1,
-1)$:
 \be\label{5sphr}%
 {\cal S}_\Lambda:  &&\eta^{}_{AB} \xi^A \xi^B= -R^2,%
\\ %
\label{ds2}%
&&ds^2=\eta^{}_{AB}\, d\xi^A \, d\xi^B , %
\ee
where $R^2:=3\Lambda^{-1}$, $A, B=0, \ldots, 4$, and $\La$ is the
cosmological constant.  Clearly, Eqs.(\ref{5sphr}) and (\ref{ds2})
are invariant under $dS$ group ${\cal G}_\Lambda =SO(1,4)$.

The $BdS$ spacetime ${\cal B}_\La$ is a $dS$ spacetime defined by
the following Beltrami coordinates on ${\cal S}_\La$ patch by
patch \cite{BdS}. Clearly, ${\cal B}_\Lambda\simeq{\cal S}_\Lambda$.
For intrinsic geometry of ${\cal B}_\Lambda$, there are at least eight
patches $U_{\pm\alpha}:= \{ \xi\in{\cal S}_\Lambda :  \xi^\alpha\gtrless 0\},
\alpha=1, \cdots, 4$. In $U_{4}$, for instance, the Beltrami coordinates are
\be \label{u4}%
&&x^\mu|_{U_{4}} =R {\xi^\mu}/{\xi^4},\qquad \mu=0,\cdots, 3;  \\
&&\xi^4=((\xi ^0)^2-\sum _{a=1}^{3}(\xi ^a)^2+ R^2 )^{1/2} > 0.
\ee
It is important that the coordinate transformation in each intersection is
a fractional linear transformation.  For example, let $y^{\mu'}$
($\mu' = 0,$ 1, 2, 4) be the coordinates on $U_3$, then, in $U_3 \cap U_4$, the transformation is $x^\mu = R\,y^\mu/y^4$ ($\mu=0,1,2$), $x^3 = R^2 / y^4$,
which is induced by
$T_{4,3}=\xi^3/\xi^4 = x^3/R = R/y^4 \in {\cal G}_\Lambda$.

In each patch, there are condition and Beltrami metric
\begin{eqnarray}\label{domain}
\sigma(x)&=&\sigma(x,x):=1-R^{-2}
\eta_{\mu\nu}x^\mu x^\nu >0,\\
\label{bhl} ds^2&=&[\eta_{\mu\nu}\sigma^{-1}(x)+ R^{-2}
\eta_{\mu\rho}\eta_{\nu\sigma}x^\rho x^\sigma
\sigma^{-2}(x)]dx^\mu dx^\nu.
\end{eqnarray}
Under fractional linear transformations of ${\cal G}_\Lambda$
\begin{equation}\label{G}
\begin{array}{l}
x^\mu\rightarrow \tilde{x}^\mu=
\sigma^{1/2}(a)\sigma^{-1}(a,x)(x^\nu-a^\nu)D_\nu^\mu,\\
D_\nu^\mu=L_\nu^\mu+{ R^{-2}}%
\eta_{\nu \rho}a^\rho a^\tau (\sigma(a)+\sigma^{1/2}(a))^{-1}L_\tau^\mu,\\
L:=(L_\nu^\mu)_{\mu,\nu=0,\cdots,3}\in SO(1,3),
\end{array}\end{equation}
where $\eta_{\mu\nu}={\rm diag}(1, -1,-1,-1)$ in $U_{\pm\alpha}$,
condition (\ref{domain}) and metric (\ref{bhl}) are invariant.
Note that Eqs.(\ref{domain})-(\ref{G}) are defined on ${\cal
B}_\Lambda$ patch by patch. This is, in fact, a cornerstone for
the special relativity-type principle. In addition, at the origin
of the coordinate system, the metric (\ref{bhl}) becomes
Minkowskian.

The generators of ${\cal G}_\Lambda$ in Beltrami coordinates are expressed as %
\begin{equation}\label{generator}
\begin{array}{l}
  \mathbf{P}_\mu =(\delta_\mu^\rho-R^{-2}x_\mu x^\rho)
  \partial_\rho, \quad
  x_\mu:=\eta_{\mu\nu}x^\nu,\\
  \mathbf{L}_{\mu\nu} = x_\mu \mathbf{P}_\nu - x_\mu \mathbf{P}_\nu
  = x_\mu \partial_\nu
  - x_\nu \partial_\mu \in so(1,3).
\end{array}
\end{equation}
Hereafter, we use bold roman letters to denote generators of Lie group.
Later we shall use a bold italic letter $\vect{x}$ to
denote the triple $(x^1, x^2, x^3)$ (see, Eq.(\ref{eml}), for example).
They form an $so(1,4)$ algebra:
\begin{eqnarray}
  [ \mathbf{P}_\mu, \mathbf{P}_\nu ] &=& R^{-2} \mathbf{L}_{\mu\nu} \nno\\
  {[} \mathbf{L}_{\mu\nu},\mathbf{P}_\rho {]} &=&
    \eta_{\nu\rho} \mathbf{P}_\mu - \eta_{\mu\rho} \mathbf{P}_\nu
\label{so(1,4)}\\
  {[} \mathbf{L}_{\mu\nu},\mathbf{L}_{\rho\sigma} {]} &=&
    \eta_{\nu \rho} \mathbf{L}_{\mu\sigma}
  - \eta_{\nu \sigma} \mathbf{L}_{\mu\rho}
  + \eta_{\mu \sigma} \mathbf{L}_{\nu \rho}
  - \eta_{\mu\rho} \mathbf{L}_{\nu \sigma}, \nno
\end{eqnarray}
as expected. The Casimir operators of ${\cal G}_\Lambda$ are (see, e.g.,
\cite{Gursey})%
\be \label{C1}
\mathbf{C}_1 := \mathbf{P}_\mu \mathbf{P}^\mu %
  -\d 1 {2R^2} \mathbf{L}_{\mu \nu} \mathbf{L}^{\mu \nu}, \qquad %
\mathbf{C}_2 := \mathbf{S}_\mu \mathbf{S}^\mu -\d 1 {R^2} \mathbf{W}^2,
\ee
where%
\be \label{sw} %
\mathbf{P}^\mu &=& \eta^{\mu \nu} \mathbf{P}_\nu, \qquad\qquad %
\mathbf{L}^{\mu \nu}=\eta^{\mu \rho} \eta^{\nu \si} \mathbf{L}_{\rho \si}, \\
\mathbf{S}_\mu &=& %
\frac 1 2 \eps_{\mu \nu \la \si} \mathbf{P}^\nu \mathbf{L}^{\la \si}, \quad %
\mathbf{S}^\mu =\eta^{\mu \nu} \mathbf{S}_\nu,\\
\mathbf{W} &=&\frac 1 8 \eps_{\mu \nu \la \si} %
  \mathbf{L}^{\mu \nu} \mathbf{L}^{\la \si}. %
\ee
In Eq.(\ref{sw}), $\eps_{\mu \nu \la \si}$ is 4-d Levi-Civita
symbol in flat spacetime with $\eps_{0123}=1$.

According to the spirit of Einstein's special relativity, one can
define simultaneity such that two events $A$ and $B$ are
simultaneous if and only if the Beltrami time coordinates $x^0$'s
for the two events coincide,
\be \label{s1}%
a^0:=x^0(A) =x^0(B)=:b^0. %
\ee
It defines the laboratory time in one patch.

The simultaneity also defines a 3+1 decomposition of spacetime
\be
ds^2 =  N^2 (dx^0)^2 - h_{ij} \left (dx^i+N^i dx^0 \right )
\left (dx^j+N^j dx^0 \right ) %
\ee
with the lapse function, shift vector, and induced 3-geometry on
3-hypersurface $\Si$ in one coordinate patch
\begin{eqnarray} \label{lapseshift}
& & N=\{\si_{\Si}(x)[1-(x^0 /R)^2]\}^{-1/2}, \nonumber \\%
& & N^i=x^0 x^i[ R^2-(x^0)^2]^{-1},
 \\
& & h_{ij}=\dl_{ij} \si_{\Si}^{-1}(x)-{ [R\si_{\Si}(x)]^{-2}
\dl_{ik} \dl_{jl}}x^k x^l ,\nonumber
\end{eqnarray}
respectively, where $\si_{\Si}(x)=\si(x)|_{x^0={\rm const.}}
=1-(x^0/R)^2 + {\dl_{ij}x^ix^j /R^2}|_{x^0={\rm const.}}$,
$\dl_{ij}$ is the Kronecker $\dl$-symbol, $i,j=1,2,3$.  In
particular, at $x^0=0$, $\si_{\Si}(x)=1+{ \dl_{ij} x^i x^j/R^2}$.

In ${\cal B}_\Lambda$, the simultaneity can also be defined with
respect to the proper time  of a clock rested at the spatial
origin of the coordinate system \cite{BdS}:
\begin{eqnarray}\label{ptime}
T_{\Lambda}=R \sinh^{-1}\left(\d{ct}{R\sigma^{1/2}(x)}\right).
\end{eqnarray}
If this proper time is chosen as the time coordinate, the metric
(\ref{bhl}) becomes the Robertson-Walker-like one as follows
\cite{BdS}:
\begin{equation}\label{dsRW}
ds^2=c^2 dT^2-\cosh^2\left(\d{cT}R\right) dl_{{\Sigma_T}}^2.
\end{equation}
where
\be\begin{array}{l}\label{spacelike}
dl_{\Sigma_T}^2:=\sigma_{\Sigma_T}^{-2}(x) \left(
\sigma_{\Sigma_T}(x) \delta_{ij} -\d 1 {R^{2}}
\delta_{ik}\delta_{jl}x^k x^l \right)dx^i dx^j, \\
\sigma_{\Sigma_T}(x,x):=1+\d 1{R^{2}}\delta_{ij}x^i x^j>0.
\end{array}\ee
The metric Eq.(\ref{dsRW}) is closely linked with the
cosmological principle.

The two kinds of simultaneity Eqs.(\ref{s1}) and (\ref{ptime})
indicate that in ${\cal B}_\La$ there is a
relation between the special relativity-type
principle, which could be introduced in the following
subsection, and the cosmological principle.


\subsection{The Inertial-type Motion and Beltrami-de Sitter Relativity
Principle \label{sect:InertBdS}}

It is important to see the advantage of Beltrami
coordinates --- both test free massive
particles and light signals in ${\cal B}_\Lambda$ are moving
along timelike straight world lines and the null ones, respectively.
In terms of straight lines, we refer to those curves with equations
in a linear form.
This indicates that both their motions
and the Beltrami coordinate systems are of the inertial-type.
Furthermore, these properties are invariant under the fractional
linear transformations (\ref{G}).

In ${\cal B}_\Lambda$ geodesics, which satisfy
\be\label{geod}%
 \frac{d^2x^\rho}{ds^2}+\Gamma^{\phantom{\mu}\rho}_{\mu\nu} \frac{dx^\mu}{ds}
 \frac{dx^\nu}{ds}=0%
\ee%
with
\be
\Gamma^{\phantom{\mu}\rho}_{\mu\nu}
  =  \d 1{R^2\sigma(x)} (x_\mu\,\delta^\rho_\nu + x_\nu\,\delta^\rho_\mu),
\label{Gamma} \ee are Lobachevski-like straight world lines
\cite{BdS, Lu, LZG, beltrami}. In particular, timelike geodesics
have the form
\begin{equation} \label{BdStlg}
x^\mu(w)=c^\mu w+b^\mu
\end{equation}
with the initial condition
\[
x^\mu|_{s=0}=b^\mu, \qquad \left . \frac{dx^\mu}{ds} \right |_{s=0}=c^\mu
\]
and the constraint
\[
g_{\mu\nu}(b)c^\mu c^\nu=1,
\]
where $w = w(s)$ is a new parameter such that
\be \label{w1} %
  w(s) =  \d {R\sinh \frac s R}
  {\frac {\eta_{\mu\nu}\,c^\mu b^\nu}{R\sigma(b)}\sinh \frac s R
    + \cosh \frac s R}.
\ee
Null geodesics have the form
\be \label{BdSnullg}
x^\mu = c^\mu w +b^\mu , \ee
with the initial condition
\[ %
x^\mu|_{\la =0}=b^\mu,  \qquad %
\left . \d {dx^\mu}{d\la }\right |_{\la =0}=c^\mu, %
\]
and the constraint
\[ g_{\mu \nu}(b)\,c^\mu c^\nu=0, \]
where $\la$ is an affine parameter and $w = w(\la)$ is a new
parameter such that
\be \label{w2}%
w(\la) = \d \la {1+\la /\la_0}
\ee
with
\begin{equation}\nonumber
\la _0=\sqrt{\frac{ R^2 \sigma (b)}{|\eta _{\mu \nu}c^\mu c^\nu |}}.
\end{equation}
Namely, free particles and light signals move along straight lines with
a uniform component velocity
\begin{equation}\label{vi}
\frac{dx^i}{dt}=v^i, \quad \frac{d^2x^i}{dt^2}=0, \quad i=1,2,3,
\end{equation}
which implies that the Beltrami coordinate system is of physical
meaning as inertial-type.

Furthermore, for a free particle 4-momentum and 4-angular momentum,
$p^\mu$ and $L^{\mu \nu}$, defined by
\be
p^\mu &:=& \d {m_{\Lambda 0}}{\sigma(x)}\frac{dx^\mu}{ds}, \\
\label{angular4}%
L^{\mu \nu}&:=&x^\mu p^\nu -x^\nu p^\mu,
\ee
are conserved along a geodesic, i.e.
\be %
\frac{dp^\mu}{ds}&=&0, \\
\frac{dL^{\mu \nu}}{ds}&=&0. %
\ee %
They constitute a 5-d angular momentum
\begin{equation}\label{angular5}
{\cal L}^{AB}:=m_{\Lambda
0}(\xi^A\frac{d\xi^B}{ds}-\xi^B\frac{d\xi^A}{ds})
\end{equation}
for a free particle in ${\cal S}_\Lambda$ in such a way that
${\cal L}^{\mu\nu} = L^{\mu\nu}$, ${\cal L}^{4\mu} = R\, p^\mu$, and it is
conserved along the geodesic,
\be \frac{d{\cal L}^{AB}}{ds}=0. \ee
The Einstein's famous formula can be generalized in ${\cal
B}_\Lambda$:
\be\label{eml}
&  -\d 1{2R^2}{\cal L}^{AB}{\cal L}_{AB}=E^2-\vect{P}^2 c^2+
\d {c^4}{R^2} \vect{K}^2-\d {c^2}{R^2} \vect{J}^2=m_{\Lambda 0}^2 c^4, &\\
& \d E c = p^0, \quad P^i=p^i, \quad K^i=\d 1 c L^{0i}, \quad
J^i=\d 1 2 \eps^{ijk}L^{jk}, & \ee
where ${\cal L}_{AB}=\eta_{AC}\eta_{BD}{\cal L}^{CD}$ and
$\epsilon^{ijk}$ is totally anti-symmetric with $\epsilon^{123}=1$. As
mentioned in the previous subsection, in this paper we use a bold
italic letter, $\vect{P}$, say, to denote the triple $(P^1, P^2,
P^3)$.  Notations such as $\vect{P}^2 = |\vect{P}|^2$,
$\vect{P}\cdot\vect{K}$ and $\vect{P}\times\vect{A}$ are
abbreviations of $(P^1)^2 + (P^2)^2 + (P^3)^2$, $P^1 K^1 + P^2 K^2
+ P^3 K^3$ and $\epsilon^{ijk}P^j K^k$, respectively. In this
sense, sometimes we do not care whether the summed indices are
one upper and one lower, as done in Newtonian mechanics.  In addition,
Eq.(\ref{eml}) gives rise the energy of photons,
\be %
E_\ga = \sqrt{\vect{P}^2c^2 -\frac {c^4} {R^2} \vect{K}^2
  +\frac {c^2}{R^2}\vect{J}^2}. %
\ee

In any case, these offer a consistent way to define the
observables for free particles and this kind of definitions differ
from any others in $dS$ space. Of course, these issues
significantly indicate that the motion of a free particle in
${\cal B}_\Lambda$ should be of inertial-type in analog to
Newton's and Einstein's conception for the inertial motion of a
free particle with constant velocity. Consequently, the coordinate
systems with Beltrami metric should be linked with inertial-type frames
and corresponding observer should be of inertial-type as well.

The deviation of nearby, `same-directed', straight world lines is
governed by the geodesic deviation equation
\be \label{gde}%
\frac {D^2 \zeta^\mu}{ds^2} + R^\mu_{~ \nu \la \si}\zeta^\la \frac
{d x^\nu}{ds} \frac {d x^\si}{ds}=0,%
\ee
where $\zeta^\mu$ is the deviation vector of nearby geodesics such
that
\be \label{dv}
g_{\mu \nu} \zeta^\mu \d {dx^\nu} {ds} =0
\ee
on the straight world lines.  For simplicity, consider two static
particles in a Beltrami coordinate system, the zero-component of
the deviation vector of which is identical to zero.  Then,
Eqs.(\ref{gde}) and (\ref{dv}) reduce to
\be
\frac {d^2 \zeta^\mu}{dt^2}
- \frac 2 {R^2}\frac{d(\si^{-1}\dl_{ij}x^i \zeta^j )}{dt} \frac {d x^\mu}{dt}=0%
\ee
and
\be
\dl_{ij}x^j \zeta^i =0.
\ee
They give rise to
\be \label{Bgde}%
\frac {d^2 \zeta^\mu}{dt^2} =0. \ee
Notice that for the two static particles $\frac {d\zeta^\mu}{dt}$ vainishes  initially.  Thus, measured by use of Beltrami coordinates, the deviation of
two static particles in the same system will be unchanged
in the evolution\cite{LZG}.  The
conclusion is not only valid in a vicinity of the spatial origin
of a system, but also valid at other places.

As was noted in \cite{BdS, Lu, LZG}, all above results indicate
that a principle similar to the principle of the special
relativity for the inertial motion with respect to the inertial
coordinates in special relativity could be set up in the
Beltrami-de Sitter spacetime ${\cal B}_\Lambda$. It is named the
Beltrami-de Sitter relativity principle.


\section{The Newton-Hooke Limit of Beltrami-$dS$ Spacetime and Inertial-type
Motion\label{NHlimit-IM}}

\subsection{The Newton-Hooke Limit of Beltrami-$dS$ Spacetime\label{NHlimit}}

In order to consider the non-relativistic limit of ${\cal
B}_\Lambda$, we set $x^0 = ct$ and $x'^0 = ct'$, and
correspondingly, we assume that $a^0 = c\,t_a$ and $b^0 = c\,t_b$
with $t_a$ and $t_b$ being finite under the limit. Naively, this
non-relativistic limit may be attained when we simply let
$c\rightarrow\infty$, but this limit is not well defined if $R$
remains finite (for algebraic reasons, see, e.g. \cite{Gibbons}).
In order to obtain the meaningful non-relativistic limit, one has
to consider the so-called Newton-Hooke limit, which is defined as
\be \label{nu}
c \to \infty, \quad R\to \infty, \quad {\rm but}\quad
  \nu = \frac c R {\rm ~ is ~ a ~ positive, ~ finite ~ constant }.
\ee%
$\nu^{-1}$ has the same dimension as time. \omits{\blue and is
called the Newton-Hooke constant.} Furthermore, $t$, $x^i$ and the
primed quantities correspondingly are assumed to be finite under
the limit.  { This is a crucial requirement for the Newton-Hooke
limit.  Otherwise, the limit will not be unique.  In
Sec.{\ref{unique-RP}}, we shall discuss the problem in details.}

Under the NH limit,
\be\label{sigma}
  \lim_{c, R\to\infty} \sigma(x) = 1 - \nu^2 t^2:=\sigma_+(t),
\qquad
  \lim_{c, R\to\infty} \sigma(a,x) = 1 - \nu^2t_a \,t:=\sigma_+(t_a,t).
\ee%
The condition $\sigma(x) > 0$ can be assured by
\be\label{t}
  - \frac{1}{\nu} < t < \frac{1}{\nu}.
\ee%

It can be shown that the transformations (\ref{G}) become
\be\label{GNH} t' = \d {t-t_a} {\sigma_+(t_a,t)}, \qquad x'^i= \d
{\si_+^{1/2}(t_a)} {\sigma_+(t_a,t)} {O^i_{~j}} (x^j - a^j -
u^jt), \qquad O^i_{~j} \in SO(3),
\ee%
which is called a \textit{Newton-Hooke transformation}.   The
first equation of Eq.(\ref{GNH}) shows that $t$ is separated from
space in the transformation and that the simultaneity for $t$ is
absolute. In this sense, we call $t$ the Newton-Hooke universal
time.  An NH transformation has inverse transformation, reading

\be
  t = \frac{t' - t_a'}{\si_+(t_a',t')}, \qquad
  x^i = \frac{\si_+^{1/2}(t_a')}{\si_+(t_a',t')}\,
  O^{~i}_j\,( x'^j - a'^j - u'^j t')
\ee
with
\be
  t_a' = - t_a, \qquad
  a'^i = - O^i_{\phantom{i}j}\,\frac{a^j + u^j t_a}{\si_+^{1/2}(t_a)}, \qquad
  u'^i = - O^i_{\phantom{i}j}\,\frac{u^j + \nu^2 t_a a^j}{\si_+^{1/2}(t_a)}.
\ee
In the inverse transformation and follows, $O^{~i}_j$ are the
entries of the inverse matrix of $O^i_{\ j}$, satisfying $O^{~i}_k
\, O^k_{\ j} = \delta^i_j.$ So the inverse transformation of an NH
transformation is still an NH transformation. Let $NH_+$ be the
set of all the NH transformations. Then it follows that $NH_+$ is
a subset of the group of diffeomorphisms. It can be further shown
that an NH transformation with parameters $t_a$, $a^i$,
$u^i$ and the rotation $O^i_{\phantom{i}j}$ followed by another
with parameters $t_a'$, $a'^i$, $u'^i$ and the rotation $O'^i_{\
\phantom{i}j}$ is equivalent to an NH transformation with the
following parameters,
\begin{eqnarray}\nno\label{eq:mult}
t_a'' = \d{t_a + t_a'}{1 + \nu^2 t_a t_a'},  \quad a''^i = a^i +
O^{\phantom{j}i}_j\,
        \d{a'^j - t_a u'^j}{\si_+^{1/2}(t_a)},  &&\\
u''^i = u^i + O^{\phantom{j}i}_j\,
        \d{u'^j - \nu^2 t_a a'^j}{\si_+^{1/2}(t_a)}, \quad
O''^i_{\ \phantom{i}j} = O'^i_{\ \phantom{i}k}
O^k_{\phantom{k}j}.&&
\end{eqnarray}
In short, the composition of two NH transformations are
still an NH transformation. Since $NH_+$, the subset of
the group of diffeomorphisms, is closed under the inverse
operation and multiplication, it follows that $NH_+$ is a
subgroup. We call $NH_+$ the Newton-Hooke group. It is a
fractional linear realization
of a  Lie group, having 10 free parameters, $t_a, a^i, u^i$, and three
parameters in $O^i_{~j}$.

Note that the transformation for time coordinate is independent of
space coordinates,
and that the time transformation is still fractional linear while the
transformation for space coordinates are linear among themselves.
In other words, if two events have the same value for coordinate $t$, they have
the same value for $t'$. Thus the simultaneity defined by the `time' coordinate
(either $t$ or $t'$) is absolute under the action of the NH group.
Such a space-time is called the Newton-Hooke space-time, denoted by
${\cal NH}_+$ as before.

{From} the transformation (\ref{GNH}), the `time-translation'
generator with respect to $t_a$ reads
\begin{equation}
{\bf H}=\sigma_+(t)\partial_t -\nu^2 t x^i\partial_i.
\end{equation}
Similarly, the `space-translation' generators with respect to $a^i$ and
boost generators
with respect to $u^i$ can be easily obtained as %
\be%
 {\bf P}_i=\partial_i, \qquad {\bf K}_i = t\partial_i, %
\ee%
respectively. The space-rotation generators are the generators
of $SO(3)$ as usual. From the above explicit forms of generators,
we can readily write down the following Lie algebra:
\be\label{eq:LieAlg}%
&[{\bf H},{\bf P}_i]=\nu^2 {\bf K}_i,\quad
[{\bf H},{\bf K}_i]= {\bf P}_i,& \nonumber \\[2mm]
&[{\bf J}_i,{\bf P}_j]=\epsilon_{ijk} {\bf P}_k,\quad [{\bf
J}_i,{\bf K}_j]=\epsilon_{ijk} {\bf K}_k,\quad
[{\bf J}_i,{\bf J}_j]=\epsilon_{ijk} {\bf J}_k& \\[2mm]
&{\rm and ~ other ~ Lie ~ brackets ~ vanish},& \nonumber
\ee%
where $\epsilon_{ijk}$ is totally anti-symmetric with $\epsilon_{123}=1$.
The Lie algebra may also be reached by taking the NH contraction of the Lie
algebra (\ref{so(1,4)}) on $BdS$ noting
\be
{\bf H}=c{\bf P}_0, \qquad {\bf K}_i=\frac 1 c {\bf L}_{0i}, \qquad %
{\bf J}_i=\d 1 2 \eps_{ijk} {\bf L}_{jk},
\ee
and it is exactly the same as $\mathrm{n}_{10}^+$ in Eq.(13) of
\cite{Gibbons} as long as setting $\nu^2=\tau^{-2}$.  The first
Casimir operator is
\be\label{eq:C1} \tilde{\bf C}_1 = {\bf P}_i {\bf P}_i-\nu^2 {\bf
K}_i {\bf K}_i, \ee
which can also be obtained from the contraction of $-{\bf
C}_1/c^2$ from Eq.(\ref{C1}). The second Casimir operator tends to
zero as $c$ (and $R$) $\to\infty$.  If the NH quantum mechanics
and other physical aspects are concerned, we must consider the
central extension of $\mathrm{n}_{10}^+$ and the corresponding
Casimir operators. We will examine these aspects in Ref. \cite{NH2}.

Obviously, the NH limit will not alter the definition of simultaneity,
Eq.(\ref{s1}).  Under the NH limit, Eq.(\ref{lapseshift}) turns out
to be
\begin{eqnarray}
& & N_+=\si_{+}^{-1}(t), \nonumber \\%
& & c N^i_+= \nu^2 t x^i \si_{+}^{-1}(t), \\
& & h_{+ij}=\si_{+}^{-1}(t) \dl_{ij} ,\nonumber
\end{eqnarray}
and the invariant Beltrami metric (\ref{bhl}) now splits to two
metrics for time and space, respectively.  They are
\be \label{tmetric} d\tau^2=\d 1 {\si_+^2(t)}dt^2= \d {dt^2}
{(1-\nu^2 t^2)^2}, \ee %
and in a hypersurface of simultaneity
\be\label{dl_3}%
d l^2=\si_{+}^{-1}(t)\,dl_0^2,\quad~d l_0^2 := \delta_{ij}\,
dx^i dx^j.%
\ee%

Note that under the NH limit the second definition of simultaneity
with respect to the proper time (\ref{ptime}) of a clock rested at
the spatial origin of the coordinate system in ${\cal B}_\Lambda$
now  coincides with the first one.
This implies that $d\tau^2$ on the whole ${\cal NH}_+$ and $d l^2$
on each hypersurface of simultaneity are defined intrisically. Namely, they
are invariant under the NH transformations.  We can examine the fact,
using Eq.(\ref{GNH}) and formulae
\begin{equation}
\si_+(t')=\d {\si_+(t_a)\si_+(t)}{\si_+^2(t_a,t)},
\end{equation}
\begin{equation}
dt'=\d {\si_+(t_a)}{\si_+^2(t_a,t)}dt,
\end{equation}
and in the hypersurface of simultaneity ($dt=dt'=0$)
\begin{equation}
dx'^i=\d {\si_+^{1/2}(t_a)}{\si_+(t_a,t)}dx^i.
\end{equation}
The fact can also be checked directly by calculating the Lie derivatives of
$d\tau^2$ and $d l^2$ with respect to all the generators.

Under the NH limit, the Robertson-Walker-like metric (\ref{dsRW}) in
${\cal B}_\Lambda$ also splits to two parts.  The detailed discussion is
left in the next section.


\subsection{Inertial-type Motion of Test Particles and Signals
in ${\cal NH}_+$ \label{motion}}

It should be noted that the space-time in the NH limit
is not a 4-d metric spacetime because there is no invariant metric on it.
In spite of that, a connection still exists as the contraction of the
connection on ${\cal B}_\Lambda$. Let the coefficients of the connection on
${\cal NH}_+$ be denoted by $\Gamma^{\ \rho}_{\mu\nu}$, where the
indices take values from $t$, $1$, $2$ and $3$. Then a direct
calculation results in all of them, among which only the following
are nonzero:
\begin{equation} \label{Cs}%
  \Gamma^{\ t}_{tt} = \frac{2\nu^2 t}{1 - \nu^2 t^2}, \qquad %
  \Gamma^{\ i}_{tj} = \Gamma^{\ i}_{jt} %
  = \frac{\nu^2 t}{1 - \nu^2 t^2}\,\delta^j_i. %
\end{equation} 
A further discussion on connection and curvature in ${\cal NH}_+$
is made in Appendix~\ref{appdx:conncurvt}.

Under the NH limit, the geodesic equation (\ref{geod}) becomes %
\be\label{geod+t}%
 \frac{d^2 t}{d\tau^2}&+&\Gamma^{\phantom{\mu}{t}}_{tt} \frac{dt}{d\tau}
 \frac{dt}{d\tau} =0,\\
\label{geod+i} \frac{d^2
x^i}{d\tau^2}&+&2\Gamma^{\phantom{\mu}i}_{{t}j} \frac{dt}{d\tau}
\frac{dx^j}{d\tau}=0.
 \ee

The general solution of Eq.(\ref{geod+t}) is
\[
t = \nu^{-1}\tanh(C_1 + C_2 \nu\tau),
\]
where $C_1$ and $C_2$ are two integral constants. Depending on
whether $dt/d\tau$ is zero or not in the initial condition, this
can be reduced to
\begin{equation}
  t = \begin{cases} \const  & \mathrm{if \quad}
  \left . \d {dt}{d\tau}\right |_{\tau=0} =0, \\
\d{1}{\nu}\,\tanh\nu\tau & \mathrm{if \quad}\left . \d
{dt}{d\tau} \right |_{\tau=0} \neq 0.
\end{cases}
\label{tau-t}
\end{equation}
For convenience, we call a curve in ${\cal NH}_+$ timelike if
$dt/d\tau$ is nonzero at every parameter $\tau$ along it, and
spacelike if it is a curve in the hypersurface $t = \const$.  A
timelike curve is also called a world line of a particle, as did
in relativity.

For a free particle with mass $m_{\Lambda 0}$, the second expression of
Eq.(\ref{tau-t}) can be directly obtained from Eq.(\ref{tmetric})
under the initial condition $\tau=0$ when $t=0$. Eq.(\ref{geod+i}) becomes %
\be %
  \frac{d^2 x^i}{d\tau^2}&+&2 \nu \tanh (\nu \tau) \frac{dx^i}{d\tau}=0, %
\ee %
which can be integrated out:
\be x^i=  {v^i}  \d{\tanh (\nu \tau)} \nu+ b^i.  \ee
This is the solution with the initial condition
\be \label{NHini}%
t|_{\tau=0}=0, \quad x^i|_{\tau=0}=b^i, \quad \left . \frac
{dt}{d\tau }\right |_{\tau=0}=1, \quad \left .
\frac{dx^i}{d\tau}\right |_{\tau=0}=v^i. \ee
By use of the second expression of Eq.(\ref{tau-t}), the geodesic appears
in an explicit form of a straight world line
\begin{equation} \label{tp}
x^i(t)=v^i t+b^i.
\end{equation}
This property is in analog with the straight line in the Beltrami
model of Lobachevski plane \cite{beltrami}.  The parameter
$t=w/c$ can also be obtained from Eq.(\ref{w1}) under the initial condition Eq.(\ref{NHini}) as long as $\eta_{\mu \nu}  c^\mu c^\nu  \neq 0$.
(The situation corresponding to $\eta_{\mu \nu}  c^\mu c^\nu  \neq 0$ does not
appear in the NH limit.)

In the NH limit, free particles
move along straight lines at constant velocities
\begin{equation}\label{nuvi}
\frac{dx^i}{dt}=v^i;\qquad \frac{d^2x^i}{dt^2}=0.
\end{equation}
Eq.(\ref{GNH}) gives rise the velocity addition law
\be v'^i := \d {dx'^i}{dt'} = \frac{O_{\;j}^i}{\si_+^{1/2}(t_a)}[\si_+(t_a,t)
v^j-u^{j}+\nu^{2}t_a (x^{j}-a^{j})]. \ee
In particular, for free particles the above expression becomes
\be \label{v+} v'^i = \d{O^i_{\;j}}
{\si_+^{1/2}(t_a)}[v^j-u^j+\nu^2 t_a(b^j-a^j)]. \ee
It is easy to see that free particles remain in uniform-velocity motions along
straight lines under time translation, spatial translation,
spatial rotation, and boost transformation.

The non-relativistic energy for a particle can be obtained in
standard way. To order $\d{\vect{v}^2}{c^2}$,
$\d{\vect{x}^2}{R^2}$, and $\d {x^i v^j} {Rc}$,
\be \si(x)=1-\nu^2 t^2+ \frac{\vect{x}^2}{R^2}, \ee
and
\be \d {dt}{d\tau} \approx \si(x)\left[1 +
\frac{\vect{v}^2}{2c^2} - \frac 1 2\left(\d{\vect{x}}{R}-\nu t
\d{\vect{v}}{c}\right)^2 \right] \ee
for a test particle. The energy and 3-momentum of a particle are
\be E=\si^{-1}(x)m_{\La 0} c^{2}\d {dt}{d\tau} \approx m_{\La
0}c^2+\frac 1 2 (1-\nu^2 t^2) m_{\La 0} \vect{v}^2 +m_{\La 0}\nu
^2 t\,\vect{x}\cdot\vect{v}- \frac 1 2 m_{\La 0}\nu^2 \vect{x}^2,
\ee
and
\be \label{Pi}
P^i =\si^{-1}(x) m_{\La 0} \d{d x^i}{d\tau}\approx m_{\La 0}v^i, \ee
respectively. The non-relativistic energy for a particle is then
\be \label{Enr} %
E_{nr} & \approx & \frac 1 {2m_{\La 0}} \vect{P}^2 -\d {\nu^2}
{2m_{\La 0}} (t \vect{P}-m_{\La 0} \vect{x})^2 = \frac 1 {2m_{\La
0}} (\vect{P}^2 -\nu^2 \vect{K}^2), %
\ee
where the second expression
in Eq.(\ref{Enr}) can be obtained directly from the generalized
Einstein formula Eq.(\ref{eml}).
Therefore, the kinetic energy in the NH limit is still
$P^2/(2m_{\La 0})$ and the non-relativistic energy for a particle
is equal to the sum
of its kinetic energy and the (negative) energy contributed from boost.  %
In particular, for a free particle, the non-relativistic energy reduces to
\be \label{Enrfree} E_{nr} \approx \frac 1 {2 m_{\La 0}}
\vect{P}^2 -\frac 1 2 m_{\La 0}{\nu ^2}\vect{b}^{2}, \ee
where $b^i$ is the initial position of the free particle in
Eq.(\ref{tp}).

As in Newtonian mechanics, time in Beltrami coordinate system is
absolute in the NH limit.  One only needs to consider the
geodesic deviation under condition $\zeta^0 =c(t-t_0)=0$.  Hence,
given a timelike geodesic $\gamma$ in ${\cal B}_\La$, we can
construct a congruence of geodesics $t =\nu^{-1}\,\tanh\nu\tau$,
$x^i = v^i(u) t + b^i(u)$.  The deviation is given by
\be
\zeta^i(t) = \frac{d v^i(u)}{du}\At{u=0}t + \frac{db^i(u)}{du}\At{u=0}
\ee
and satisfies
\be \label{NHBde} %
\frac {d^2\zeta^i}{dt^2} = 0 %
\ee
obviously. Eq.(\ref{NHBde}) can also be obtained from the contraction of
Eq.(\ref{gde}).   It is
invariant from one inertial-type frame to
another inertial-type frame.

Similarly, a light signal moves globally along a null geodesic in
${\cal B}_\Lambda$. The null geodesic equations formally still have the forms
of Eqs.(\ref{geod+t}) and (\ref{geod+i}),
but now \be \d {dt} {d\la} =0, \ee where $\la $ is an affine
parameter. Therefore, a null geodesic in ${\cal B}_\Lambda$ becomes a
spacelike geodesic in ${\cal NH}_+$, which can be integrated as a straight line
\be \label{NHnullg}
 x^i = c^i \la +b^i
\ee
from Eq.(\ref{geod+i}) under the initial condition
\be %
t|_{\la =0}=0, \quad x^i|_{\la =0}=b^i,  \quad \left . \d {dt}
{d\la } \right |_{\la =0}=0, \quad \left . \d {dx^i}{d\la } \right
|_{\la =0}=c^i. \nno %
\ee
Eq.(\ref{NHnullg}) can be obtained from Eq.(\ref{BdSnullg}) )
in the limit of $\la_0 \to \infty$ in Eq.(\ref{w2}). (Again, the situation for finite $\la_0$ in Eq.(2.25) does not appear in the NH limit.) Now, the geodesic deviation becomes
\be %
\frac{d^2\zeta^i}{d\la^2} = 0. %
\ee
%


\section{On Uniqueness of the Newton-Hooke Limit, Principles of Galilei-Hooke's
Relativity and Postulate on Newton-Hooke Universal Time\label{unique-RP}}


\subsection{On Uniqueness of the Newton-Hooke Limit \label{unique}}

Usually, the NH limit is defined by both $c$ and $R$
$\to \infty$, but $c/R$ keeping fixed \cite{Bacry}. It should be
noted, however, that such a  definition of limit is not well
defined because this contraction depends on the realization of the
$dS$ group acting on the $dS$ spacetime. Even for the same kinds
of coordinates, it also depends on whether the spatial coordinates
are dimensional or dimensionless.

The followings are several possible NH limits of de
Sitter spacetime.

It has been shown that the $BdS$ spacetime, based on the
definition of proper-time simultaneity, has the form in
(\ref{dsRW}). Under the NH limit, the geometry becomes
\be %
d\tau^2= dT^2,
\ee
and
\be
d l^2 = \cosh^2(\nu T)dl^2_{\Si_T}.
\ee
In this case, time spans a 1-d Euclidean metric space while the
spatial space is conformal to the 3-d Euclidean metric space.  The
conformal factor depends on time $T$. As was noted previously, in
${\cal B}_\La$  spacetime we can define two different kinds of
simultaneity. In the NH limit, however, the two kinds  of
definition of simultaneity coincide as was mentioned in the
previous section because of Eq.(\ref{tmetric}).

If one introduces dimensionless spatial Beltrami coordinates such
that
\be
\tilde x^i= \d 1 R x^i,
\ee
then
\be
\si_\Si =1 - x^ix^i/R^2=1-\tilde x^i \tilde x^i
\ee
and the Robertson-Walker-like metric can be rewritten as
\be
ds^2 =c^2 dT^2-R^2\cosh^2(cT/R)d\tilde l^2,
\ee
where $d\tilde l^2$ is the line-element on the unit 3-sphere,
which is dimensionless. Now, taking the NH limit, in
which the dimensionless coordinates $\tilde{x}^i$ are kept finite,
we get
\be \label{nrdS} d\tau^2= dT^2- \d 1 {\nu^2} \cosh^2 (\nu
T)d\tilde l ^2. \ee 
In this case, the concept of spacetime and de
Sitter group as well remain even in the NH limit!
Namely, the `contraction' is trivial in the
group, algebra and geometry aspects! It is also the case for the
Beltrami metric (\ref{bhl}) if the spatial coordinates chosen
as of dimensionless.

Similar uniqueness problem also appears in other forms of de
Sitter spacetime. For example, for $k=0$ de Sitter metric,
\be \label{NHk0}
&&ds^2 = c^2 d\bar t^2 -  e^{2c\bar t/R}(d\bar r^2+ \bar r^2 d\Om^2)
\nonumber \\
&&\Longrightarrow \begin{cases}
d\tau^2=d\bar t^2, \qquad d l^2 = e^{2\nu \bar t}(d \bar r^2+ \bar r^2 d\Om^2),   & \bar r ~ {\rm keeps ~
finite.}   \\
d\tau^2 = d\bar t^2 - \d 1 {\nu^2} e^{2\nu \bar t}(d\tilde {\bar r}^2 + \tilde {\bar r} ^2 d\Om^2), &
\tilde {\bar r} =\bar r/R ~ {\rm keeps ~ finite.} \end{cases}
\ee
For the static de Sitter metric,
 \be \label{SdS}
&&ds^2 = (1-{r'^2}/{R^2})c^2 dt'^2 - \d {dr'^2} {1-{r'^2}/{R^2}} -r'^2 d\Om^2  \nonumber \\
&&\Longrightarrow \begin{cases}
d\tau^2=dt'^2, \qquad dl^2 =dr'^2+r'^2 d\Om^2,   & r' ~ {\rm keeps ~
finite.}   \\
 d \tau ^2  = (1-\tilde {r}'^2)d{t'}^2-\d 1 {\nu^2} \left (\d
{d\tilde {r}'^2}{1-\tilde {r}'^2}+ \tilde {r}'^2
d\Om^2\right ), & \tilde r' =r'/R ~ {\rm keeps ~ finite.}
\end{cases}
\ee

If one requires $\tilde t=t/R$, and $x$ keep finite or requires
one or two of spatial coordinates $\tilde x^i=x^i/R $ keep finite,
then one may get different contractions of the same geometry (See,
appendix \ref{appdx:limit}).

It should be noted that the starting point of Ref.\cite{Gibbons}
is the static de Sitter metric.  Under the NH limit in the
first manner in Eq.(\ref{SdS}), $t'$ is equal to the proper time
and the spatial coordinates $q^i$ satisfy $q^i q^i=r'^2$.  When
such a coordinate system is chosen, free particles in the NH limit
of $dS$ spacetime obeys
\be \label{GNHgeod} %
\d {d^2 q^i}{d\tau^2} = \nu^2 q^i . \ee
It can be proved that it is the contraction of geodesic equation
in the static $dS$ spacetime, too \cite{Gibbons}. From
Eq.(\ref{GNHgeod}), one cannot read out that free particles move
along straight lines at constant velocities.  The relation between
coordinates used in \cite{Gibbons} and the Beltrami
coordinates is given by the second expression of
Eq.(\ref{tau-t}) or %
\be\nonumber%
\tau =\d 1 {\nu} \tanh^{-1} \nu t,%
\ee%
 and
\be
q^i= \d {x^i} {\sqrt{1-\nu^2 t^2}}.
\ee%
It should be noted that in terms of the Beltrami
coordinates $(t, x^i)$ Eq.(\ref{GNHgeod}) turns to be
Eq.(\ref{nuvi}) of the inertial-type motion  and the
non-relativistic energy of a particle, Eq.(5) in Ref.
\cite{Gibbons}, turns to be Eq.(\ref{Enr}).

These examples show that the condition Eq.(\ref{nu}) is not enough
to determine the limit. Is it possible that there exist one or
more physical principles to determine the NH limit uniquely? The
answer is yes. In the following subsections, we propose two
principles, Galilei-Hooke's relativity principle and
postulate on Newton-Hooke universal time $t, |t|\leq
\nu^{-1}=R/c$, and show that these two principles should fix the
NH limit procedure.


\subsection{{Principle of Galilei-Hooke's Relativity } \label{GHRP}}

Recall that in Newtonian mechanics, the Galilean relativity
principle is respected. The principle states that the laws of
mechanics are the same in all inertial frames. In particular, the
first law of Newtonian mechanics, which says that a body at
rest or in uniform-velocity motion along a straight line remains
its state as long as no force acts upon it, is the same in all inertial
frames. Therefore, an inertial observer cannot determine by local
mechanic experiments whether he is at rest or in a uniform-velocity motion
along a straight line.

In the previous discussion on the NH limit, the relativity
principle is not considered even though the inertial-type motion
has been introduced in both $BdS$ spacetime and NH space-time. We
have seen that if in ${\cal NH}_+$ and the constant curvature spacetime
resulted from the trivial contraction in previous subsection, the
inertial-type motion, the motion at uniform-velocity along
straight line in Sec. \ref{motion}, is regarded as a genuine
inertial motion, if a Beltrami coordinate system is regarded as a
true inertial coordinate system, and if a set of particles static
in a Beltrami coordinate system is regarded as an inertial
reference frame,
then the inertial motion is independent of the choice of inertial
frames.  The statement may serve
as the counterpart of the first law of Newtonian mechanics in
${\cal NH}_+$ and can be expressed in the same
way as the first Newton's law of mechanics: in ${\cal NH}_+$
a body at rest or in uniform-velocity motion along a
straight line remains its state as long as no force acts upon it.
The counterpart of the second Newton's law of
mechanics may also be set up
in ${\cal NH}_+$ as
\be \label{2ndlaw} \d{d P^i}{dt}=F^i. \ee
In ${\cal NH}_+$, both sides of Eq.(\ref{2ndlaw}) should
transform in the same manner under the transformation Eq.(\ref{GNH}).
Now, we may introduce a relativity principle parallel to the
Galilean relativity principle in Newtonian mechanics as follows:

{\it The laws of mechanics in the Newton-Hooke space-time are the same in all
inertial frames.}\\
We name it Galilei-Hooke's relativity principle in honor of
Galilei and Hooke.  (From now on, the `-type' in phrases, such as,
inertial-type motion, inertial-type observer, inertial-type frame,
are removed.)

By use of Galilei-Hooke's relativity principle, the contraction
like the third example in Appendix \ref{appdx:limit} can be excluded.
Furthermore, the limit that $c \to \infty$ while $R$ keeps finite can also be
excluded by the Galilei-Hooke's relativity principle. Therefore, the
requirement $R \to \infty$ in the non-relativistic approximation of
special-like relativity on ${\cal B}_\La$ \cite{BdS} can be relaxed
because it is the result of the Galilei-Hooke's relativity principle.


\subsection{Postulate on Newton-Hooke  Universal Time \label{ISVO}}

It should be noted that even when the Galilei-Hooke's relativity
principle is respected, it is still not enough to fix
the NH limit uniquely.

In order to fix the contraction procedure, it is needed to explore
what should happen at any given point as the origin of a
coordinate system chosen. In the case of $BdS$, the metric at
origin is Minkowskian and under the NH contraction, it splits to
two parts. The similar situation also happens in the case of
static $dS$ metric. Since both spacetime ${\cal B}_\La$ and
space-time ${\cal NH}_+$ are homogeneous  and isotropic, all
points should follow this behavior.  Note that for the metric at
the origin, there is no room for $R$. Thus, this may fix the
contraction procedure. This may also be seen from both the
light-cone and the homogeneous Lorentz group at the origin.  In
fact, we should define the inertial motion, inertial coordinates
as well as inertial observers and require two first principles for
$BdS$: the relativity principle and the postulate on
universal constants, which require that there exist two universal
constants of $c$ with dimension of velocity and $R$ with dimension
of length.  As was indicated in \cite{BdS}, the second
postulate implies that the light cone at the origin is the same as
that in a Minkowski spacetime.  The lack of the two principles
seems to indicate in mathematics that the contraction procedure
may not be fixed by only considering algebraic contraction.

Therefore, we propose the second physical principle, the postulate on the
Newton-Hooke universal time:

{\it There exits a bounded Newton-Hooke universal time $t, |t| <
\nu^{-1}=R/c$ under the NH limit of $c, R \to \infty${, which
measures the time of physical processes}.}

\bigskip

Thus, under the postulate, especially  at origin of the (Beltrami) coordinates,
the metric splits to two parts under the limit of $c \to
\infty$. In brief, only considering algebraic contraction is not
enough to uniquely determine the NH limit. One has to supplement
two physical principles. One is the Galilei-Hooke's relativity
principle and the other is the postulate on Newton-Hooke universal time.

\section{Further Study on Inertial Reference Frames}
\label{sect:GRef}

\subsection{Relative Velocity, Inertial Frame and Observers}

In Newtonian mechanics and special relativity, if two particles have the same
velocity relative to one inertial frame, they have the same velocity relative
to every inertial frame.  As a consequense, there cannot
be a particle that is at rest at the same time in two
different inertial frames.
In addition, in Newtonian mechanics and special relativity, any two
inertial frames $S$ and $S'$ have the property that all
particles being at rest in $S'$ have the same velocity relative to
$S$. In the
following, we shall see these are not the facts in ${\cal NH}_+$.

Similar to what in Newtonian mechanics and special relativity, the
equation of motion for a free particle is Eq.(\ref{tp}).
Especially, when $t_a = 0$, two  coordinate systems for $S$ and
$S'$ related by the transformation (\ref{GNH}) have the same
relation as that in Newtonian mechanics, except the coordinate
time $t$ taking value in $(-\nu^{-1},\nu^{-1})$ in ${\cal NH}_+$.
When $t_a \neq 0$, however, the velocities of free particles
at rest in $S'$ are different from one another in $S$.  This can be
seen from the equation of motion of a static particle in $S'$ in terms of $S$,
\begin{equation}
  x^i = a^i + \sigma_+^{-1/2}(t_a)\,O^{\ i}_j x'^j
  + [u^i - \nu^2 t_a\,\sigma_+^{-1/2}(t_a)\,O^{\ i}_j x'^j]\,t.
\end{equation}
Obviously, when $t_a \neq 0$, there is a special fixed spatial point
\begin{equation}
 x'^i= \frac{\sigma_+^{1/2}(t_a)}{\nu^2 t_a}\,O^i_{\ j} u^j
\end{equation}
in $S'$, whose velocity relative to $S$ is also zero!  Namely, its spatial
coordinates in  $S$ is also fixed at
\begin{equation}
  x^i = a^i + \frac{u^i}{\nu^2 t_a}.
\end{equation}

Therefore, when $t_a \neq 0$, there is always a free particle $P_0$
at rest in both frames $S$ and $S'$. Let us select another two distinct
particles $P$ and $P'$ such that $P$ is static in $S$ while $P'$ is static
in $S'$. From the point of view of an observer in $S$, $P_0$ and $P$ have
the same velocity while $P_0$ and $P'$ have different velocities. From the
point of view of an observer in $S'$, it is on the contrary: $P_0$ and $P'$
have the same velocity while $P_0$ and $P$ have different velocities. So,
in this case, the observation that two particles have the same velocity is
a phenomenon depending on reference frames: Two frames could share a common
particle that is static in both of them. For such a particle it could say that
another remote particle is static relative to it while, at the same time, it
could deny this statement. The answer depends on which reference frame it
thinks itself belongs to.

Customarily, we use an inertial observer to take the place of an
inertial frame in which he is static. This is due to the
properties of inertial frames in Newtonian mechanics and special
relativity. From the above discussion we find that this is
misleading in ${\cal NH}_+$, because the same
inertial observer can be static in distinct inertial frames. In
fact, as we can see more clearly in next section, the choice of
the time origin of NH inertial  coordinate systems has much more
nontrivial meaning than the Newtonian case. Especially, the
concept of static state is dependent on this choice.

At last, we point out that, whenever $t_a \neq 0$, we can shift the
spatial origins of $S$ and $S'$, respectively, so that the transformation from
$S$ to $S'$ have the standard form as
\begin{equation}\label{eq:std}
  t' = \frac{t - t_a}{\sigma_+(t_a, t)}, \qquad
  x'^i = \frac{\sigma_+^{1/2}(t_a)}{\sigma_+(t_a,t)}\,O^i_{\ j} x^j.
\end{equation}


\subsection{On Euclid's Fifth Axiom}

In this subsection we try to give a clear geometric picture to the NH
transformations. We can see the shadow of projective geometry.

In ${\cal NH}_+$ a 2-d surface
\begin{equation} \label{tlp}
  t = u^0, \qquad x^i = V^i u^0 + A^i u^1 + a^i
  \quad (\vect{V}^2 + \vect{A}^2 \neq 0)
\end{equation}
is called a timelike plane and a 2-d surface
\begin{equation} \label{slp}
  t = \const, \qquad x^i = A^i_0 u^0 + A^i_1 u^1 + a^i
  \quad (\vect{A}_0\times\vect{A}_1 \neq 0)
\end{equation}
is called a spacelike plane, where $u^0$ and $u^1$ are parameters
describing the surfaces and $V^i$, $A^{i}$'s and $a^i$ are some constants.
Since Eqs.(\ref{tlp}) and (\ref{slp})
remain the same forms, respectively, under the NH
transformations, the above concepts are well defined.

It can be shown that straight lines on the planes in ${\cal NH}_+$,
which can be expressed by
\begin{equation} \label{lp}
  k_0 u^0 + k_1 u^1 = l \qquad
\end{equation}
in terms of the parameters $u^0$ and $u^1$,
are also straight lines (i.e. geodesics) in ${\cal NH}_+$.  In Eq.(\ref{lp}),
$k_0$, $k_1$ and $l$ are some constants, and $u^0$ takes value in
$(-\nu^{-1}, \nu^{-1} )$ for timelike planes and $(- \infty , \infty )$ for spacelike planes.  Two geodesics in a plane are called
parallel lines if they have no common points.
It is obviously that Euclid's fifth axiom is
valid on spacelike planes but invalid on timelike planes.
Applying the same ideas to discuss spacetimes of
Newtonian mechanics or special relativity, we can find that
Euclid's fifth axiom is valid for every plane. This is one of the
main difference between ${\cal NH}_+$ and the
spacetime in Newtonian mechanics or the special relativity.

The concept of straight line, plane and parallel are independent of
the choice of Beltrami coordinate systems. That is, if a curve is a
straight line in one coordinate system, it is a straight line in all
coordinate systems; If two straight lines are parallel in one coordinate
system, they are parallel in all coordinate systems. For an inertial frame
$S$, the world lines $\vect{x} = \vect{x}_1$ and $\vect{x} = \vect{x}_2$ are
straight lines that are parallel to each other. From the point of view other
inertial frame, $S'$, say, these world lines are still parallel straight lines.
It does represent the invalidity of Euclid's fifth axiom that there exists
an observer who can sit statically in both $S$ and $S'$.

\begin{figure}
\includegraphics[width=100mm,height=80mm]{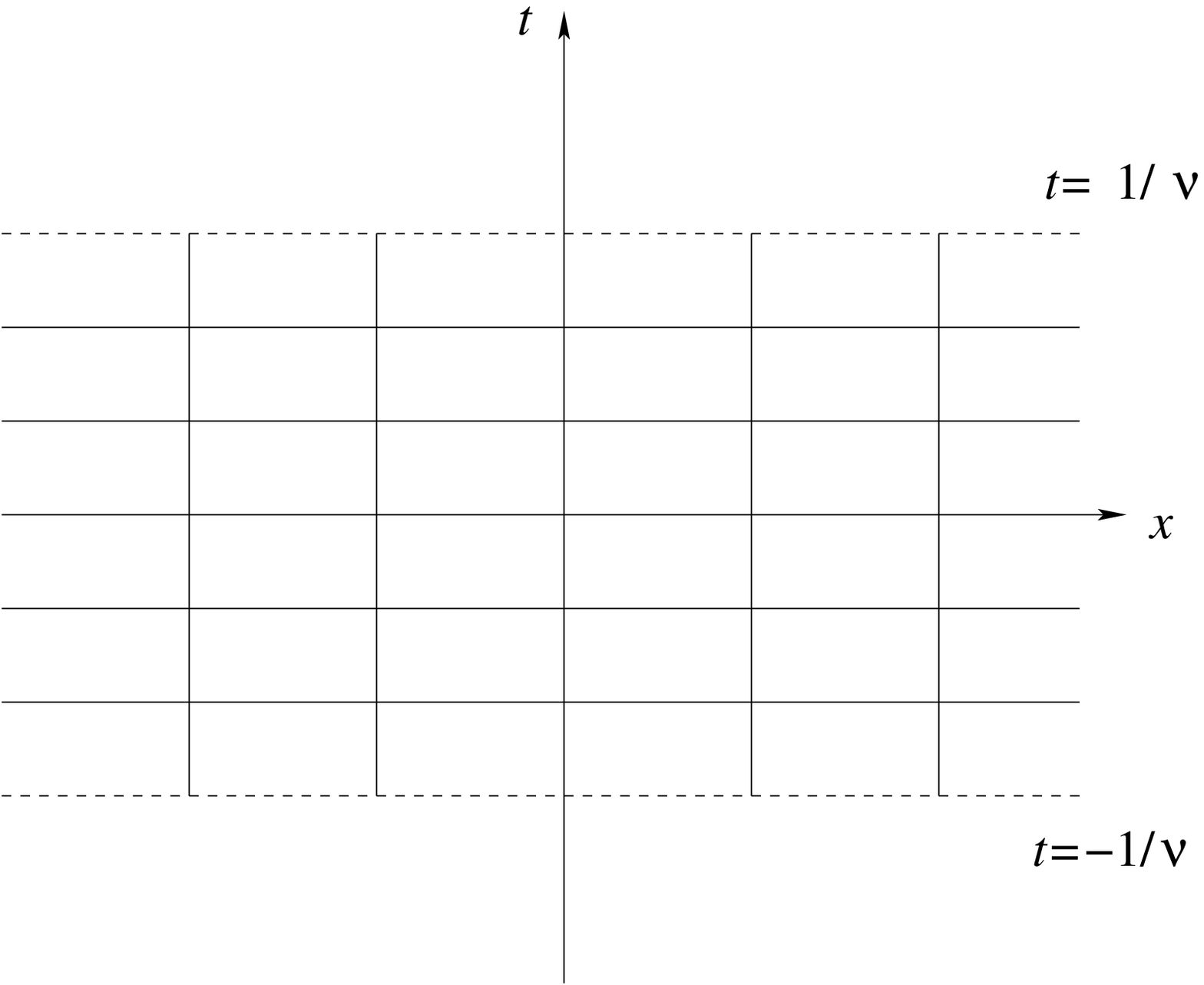}
\caption{World lines of fixed points $x=\const$.
  Horizontal lines are $t = \const$.}
\label{fig:S}
\end{figure}

\begin{figure}
\includegraphics[width=100mm,height=80mm]{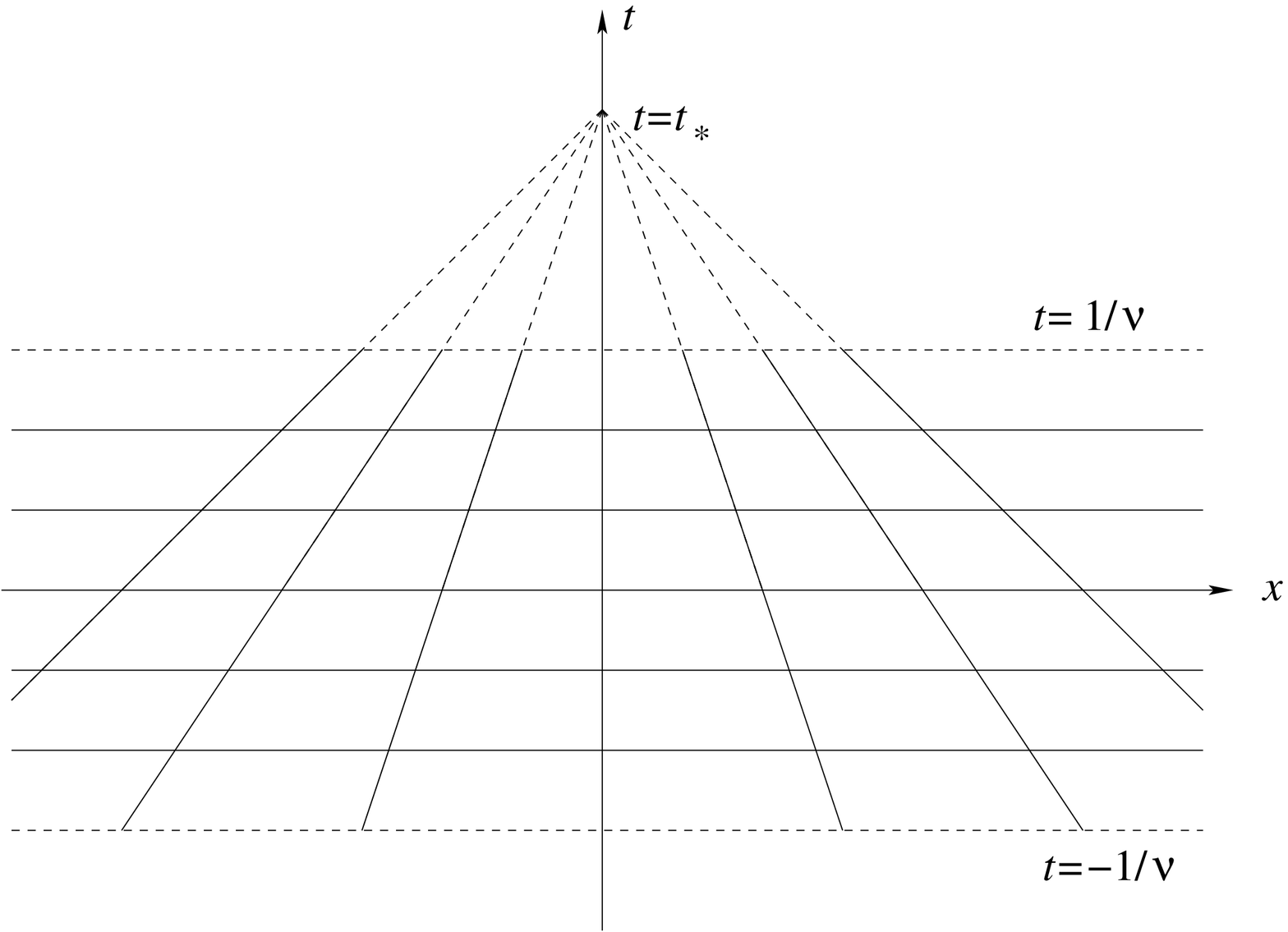}
\caption{World lines of fixed points $x' = \const$.
  Horizontal lines are $t = \const.$}
\label{fig:Sprime}
\end{figure}

To be clearer, we give some diagrams to the standard transformation
(\ref{eq:std}). Define
\begin{equation}
 t_* = \frac{1}{\nu^2 t_a}.
\label{eq:dualt}
\end{equation}
It is meaningful for all nonzero $t_a$ and can be generalized to
the case $t_a=0$, that can be regarded as a special case in which
$t_*$ is at infinity on the time axis.  It should be noted that
$t_*$ is not a possible coordinate time because $|\nu t_* | > 1.$
For simplicity we only consider a plane $\Sigma$ containing the
world line of the spatial origin of the systems, namely, $\vect{x}
= 0$ and $\vect{x}' = 0.$ If the world lines of $x = \const$ are
shown as in Fig.\ref{fig:S}, the world lines of $x' = \const$ will
be shown, in the same coordinate system, as in Fig.\ref{fig:Sprime}.
In this figure, these world lines
are parallel to each other and are focused on the point $t = t_*$
on the time axis. Conversely if a congruence of geodesics
is focused on a point in $|t| > \nu^{-1}$ (not necessarily on the time axis),
we can say that they are the world lines of rest particles in certain an
inertial frame. Due to the nature of fractional linear transformations, the
cases in Fig.\ref{fig:S} and Fig.\ref{fig:Sprime} are
symmetric{\darkgreen .} If we draw the world lines in the $x'$-$t'$ coordinate system, then
the world lines $x' = \const$ will look like those in Fig.\ref{fig:S}, while
the world lines $x = \const$ will be focused on $t' = - t_*$.
\begin{figure}
\includegraphics[width=100mm,height=80mm]{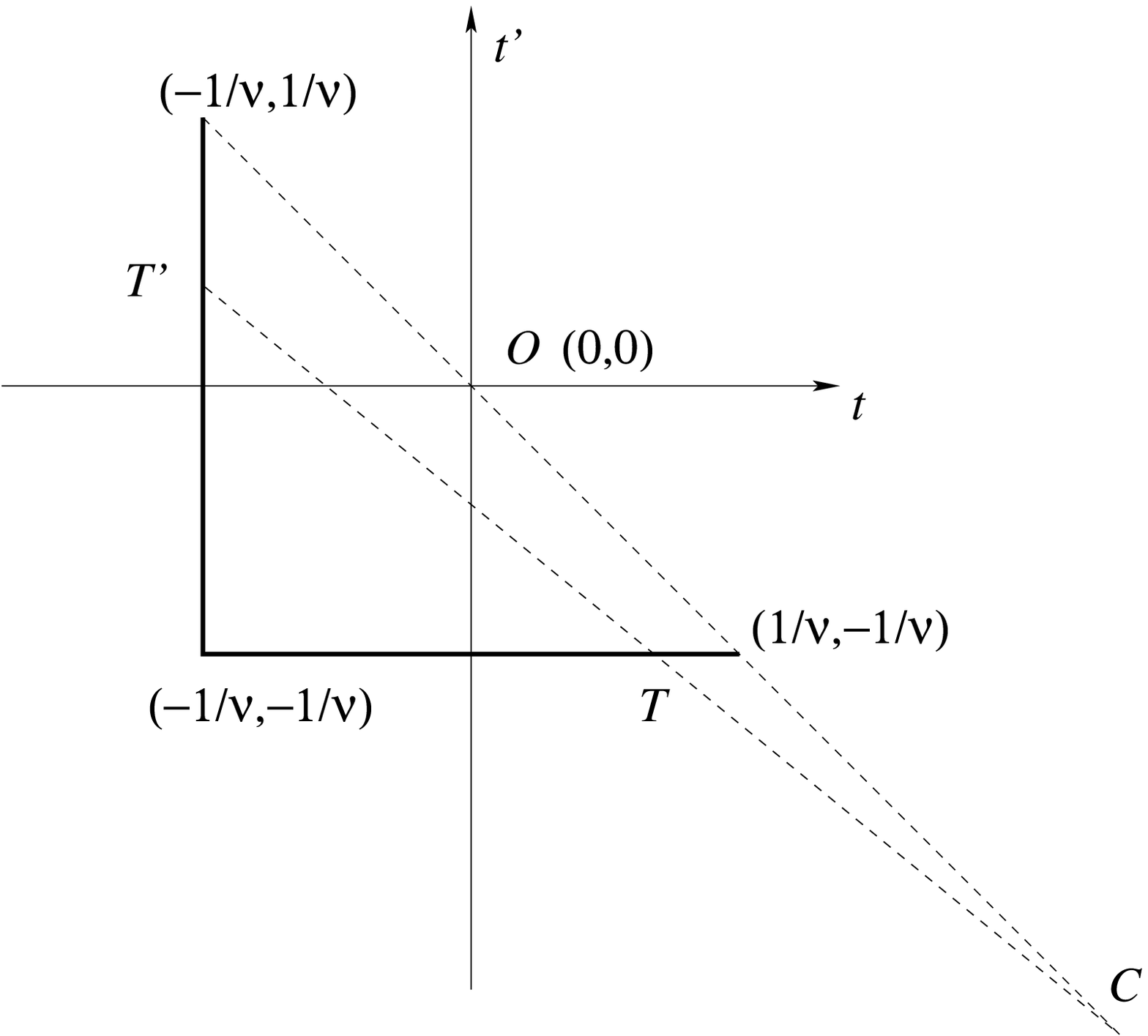}
\caption{Diagram for transformation of coordinate time.}
\label{fig:t}
\end{figure}

Fig.\ref{fig:t} is a graphic illustration for the transformation
of coordinate time. In the diagram, the horizontal $t$-axis and
the vertical $t'$-axis form a Cartesian coordinate system. The
point $C$ is located at $(t_*, -t_*)$ where $t_*$ is given in
Eq.(\ref{eq:dualt}). Choose points $T$ and $T'$ such that they are
collinear with $C$. Let the coordinates of $T$ and $T'$ be
$(t,-1/\nu)$ and $(-1/\nu, t')$, respectively.
Then, $t$ and $t'$ can be linked by the transformation
(\ref{eq:std}). From Eq.(\ref{eq:std}) and the diagram, we can see
$\pm 1/\nu$ are invariant time in ${\cal NH}_+$.  Compared with
special relativity in which there exists an invariant velocity
$c$, the NH mechanics is a kind of relativity due to the
existence of finite extremum time.

\section{Concluding Remarks}

We have shown that in ${\cal B}_\Lambda$, the transformation group
between Beltrami coordinate systems contracts to $NH_+$ group in
the NH limit. The geometry of $BdS$ spacetime splits time and
spatial space in the NH limit. The former may be regarded as a
subspace of $\mathbb{R}P^1$ with metric\footnote{ The cross ratio
is an invariant and the metric is its infinitesimal form.}, while
the latter is conformal to a 3-d Euclidean space. The conformal
factor depends on time only. Test particles and light signals
without force acting upon them move along straight lines at
uniform velocity in the NH limit, as expected. The behavior of
uniform-velocity motion along straight lines is independent of the
inertial frame which observer is in and the velocity addition law
is obtained. This can be regarded as the key point of
Galilei-Hooke's relativity principle, which together with the
postulate on  Newton-Hooke universal time determines the NH limit uniquely.

The contraction of $BdS$ spacetime shows that the Beltrami
coordinate systems should be regarded as inertial coordinate system both
in relativistic and non-relativistic levels since test particles
and signals without force acting upon them all move along straight
lines at uniform velocities.  This implies that the mechanics in
non-zero constant-curvature space-time may be set up
in parallel to the one in flat space-time, starting from the
relativity principle and inertial law.  The dynamics both at
classical and quantum level will be studied elsewhere.

It is remarkable that unlike in Newtonian mechanics and special relativity,
a free particle in ${\cal NH}_+$ (and ${\cal B}_\La$) can be static at the
same time with respect to two different inertial frames which have relative
velocity. This property is closely related to the violation of Euclid's fifth
axiom. In this sense, the NH limit of $BdS$ spacetime can be regarded as the
version of the Newtonian mechanics without Euclid's fifth axiom.

Another remarkable property of ${\cal NH}_+$ is that there is a
universal  time.  It is similar to the Newton's universal time in Newtonian
mechanics in the sense that it is separated from space in the
metric and that the simultaneity is absolute. However, in any
inertial coordinate system, this universal time as the coordinate
time should be in the range $(-1/ \nu, 1 / \nu)$.  Compared with
special relativity in which there exists an invariant velocity
$c$, the NH mechanics looks like a kind of relativity due to the
existence of finite extremum universal time. It is related to the
fact that the generators of spatial translation and boost along
the same direction form an $SO(1,1)$ vector under the time
translation. [See, Eqs.(\ref{eq:mult}), (\ref{eq:LieAlg}), and
(\ref{eq:C1}).] Further studies on space-time, group structures,
and kinematics and
dynamics in the NH limit are needed.%

It should be noted that when the NH limit in ${\cal B}_\Lambda$
is taken, the zeroth coordinate $\xi^0 \to \infty $. In order
Eq.(\ref{5sphr}) to be meaningful, one of $\xi^a \ (a=1,2,3,4)$,
at least, should tend to $\infty$. For example, in $U_4$, $\xi^4
\to \infty$ while $\xi^i \ (i=1,2,3)$ keep finite so that $x^i <
\infty$ from Eq.(\ref{u4}). The intersection of $U_3$ and $U_4$,
$U_4 \bigcap U_3$, can only appears at $\xi^3, \xi^4$ (and thus
$x^3, x^4$) $\to \infty$.

When $\nu =0$, namely, $R \to \infty$ faster than $c\to \infty$,
all results become the correspondences in flat space-time. In
particular, the Galilei-Hooke's relativity principle reduces to
the Galilean one. Non-relativistic energy of a test particle
reduces to the one in the Newtonian mechanics. The NH group
reduces to the Galilei group.  The Newton-Hooke universal
time becomes Newton's universal time.

In the present paper, we only discuss the contraction of $BdS$
spacetime which has a positive cosmological constant. All results
are readily extended to the contraction of Beltrami-anti-de Sitter
spacetime which has negative cosmological constant.


\appendix

\section{On Connection and Curvature}
\label{appdx:conncurvt}

As pointed out in Sec. \ref{sect:BdS} and in \cite{BdS,Lu,LZG}, in
the Beltrami-de Sitter spacetime ${\cal B}_\Lambda$
the metric %
\begin{equation}\nonumber
  g_{\mu\nu} = \eta_{\mu\nu}\sigma^{-1}(x)
  + R^{-2} \eta_{\mu\rho}\eta_{\nu\sigma}x^\rho x^\sigma \sigma^{-2}(x)
\label{g}
\end{equation}
and its inverse
\begin{equation}
  g^{\mu\nu} = \sigma(x)(\eta^{\mu\nu} - R^{-2} x^\mu x^\nu)
\end{equation}
are invariant under the fractional linear transformation (\ref{G}).
Consequently, the connection coefficients Eq.(\ref{Gamma}) and the components
of the curvature tensor
\begin{equation} \label{cc}
  R^\mu_{\nu\rho\sigma} = R^{-2}\,(g_{\nu\rho}\delta^\mu_\sigma
    - g_{\nu\sigma} \delta^\mu_\rho) {\yellow ,}
\end{equation}
have the same property.  Since the
coordinate transformation from one patch to another is a special
case of the fractional linear transformation (\ref{G}), the forms
of these quantities are the same in all the coordinate patches and
the expressions of the geodesics have the same form in all the patches, too.

The connection on ${\cal B}_\Lambda$ is torsion free.
The connection on ${\cal NH}_+$ is torsion free, too.  Since not a particular
Beltrami coordinate system is specified
in obtaining Eq.(\ref{Cs}) from the contraction of coefficients of connection,
Eq.(\ref{Cs}) should remain the same form under the
NH transformations.  In fact, we can
define a new set of symbols $\Gamma'^{\ \rho}_{\mu\nu}$ among which only the
following are nonzero:
\begin{displaymath} %
  \Gamma'^{\ t}_{tt} = \frac{2\nu^2 t'}{1 - \nu^2 t'^2}, \qquad %
  \Gamma'^{\ i}_{tj} = \Gamma'^{\ i}_{jt} %
  = \frac{\nu^2 t'}{1 - \nu^2 t'^2}\,\delta^j_i. %
\end{displaymath} %
Then it can be verified that $\Gamma^{\ \rho}_{\mu\nu}$ and
$\Gamma'^{\ \rho}_{\mu\nu}$ satisfy the standard transformation
relation for coefficients of connection if
$(t,\vect{x}) \rightarrow (t',\vect{x}')$ is an NH transformation.
So, the connection in ${\cal NH}_+$ is well defined.

The nonzero components of the curvature tensor and Ricci tensor calculated
from Eq.(\ref{Cs}) are, respectively
\begin{equation}
  R^i_{t\mu\nu} = \frac{\nu^2}{(1 - \nu^2 t^2)^2}\,
  (\delta^t_\mu \delta^i_\nu - \delta^i_\mu \delta^t_\nu)
\quad\textrm{and}\quad
  R_{tt} = - \frac{3\,\nu^2}{(1 - \nu^2 t^2)^2}.
\end{equation}
It follows that the mathematical forms of the curvature tensor and Ricci
tensor are also invariant under the NH transformation.  And, on the
other hand, they can also be obtained by contracting the curvature tensor and
Ricci tensor on ${\cal B}_\Lambda$, respectively.

Finally, we re-study the deviation equation of geodesics in
${\cal NH}_+$ in terms of 4-d language.

Given a timelike geodesic $\gamma$ in ${\cal NH}_+$, we can construct a family
of geodesics,
$t = \nu^{-1}\tanh^{-1}\nu\tau,$ $x^i = x^i(\tau,u)$,
labeled by the parameter $u$.
For convenience, it will be denoted by $x^\mu = x^\mu(\tau, u)$
with $\mu = t$, $1$, $2$ and $3$ and $x^t := t$.
Assume that the geodesic labeled by $u = 0$ is $\gamma$. Then, for each geodesic,
Eqs.(\ref{geod+t}) and (\ref{geod+i}) are equivilent to
\begin{equation} %
  \frac{{\r }^2 x^\mu}{{\r} \tau^2} %
 + \Gamma^{\ \mu}_{\rho\sigma}\frac{\partial x^\rho}{\partial\tau}
    \frac{\partial x^\sigma}{\partial\tau} = 0,
\label{eq:gfam}
\end{equation} %
where $\frac {\r x} {\r \tau}:= \frac {\r x(\tau, u)} {\r \tau}$, etc.
The deviation $\zeta^\mu$ is a vector field along $\gamma$, defined by %
\begin{equation} %
  \zeta^{\mu}(\tau) = \frac{\partial x^{\mu}}{\partial u}\At{u=0}.
\end{equation} %
Obviously, $\zeta^t = 0$.
If we take the partial derivative with respect to $u$ on both sides of
Eq.(\ref{eq:gfam}) then set $u = 0$, we can obtain the
derivation equation
\begin{equation}
  \frac{D^2\zeta^{\mu}}{d\tau^2} %
 + R^{\mu}_{\nu\rho\sigma}\dot{\gamma}^\nu\zeta^\rho\dot{\gamma}^\sigma = 0,
\label{eq:der} %
\end{equation}
where $\dot{\gamma}^\nu$ is the tangent vector of $\gamma$.

Using the expression of connection coefficients and the fact that $\zeta^t = 0$,
we can verify that the $t$-component of the covariant derivative of
$\zeta^\mu$ is always zero. It follows that, when $\mu = t$, Eq.(\ref{eq:der})
is an identity. When $\mu = i$, it is equivalent to
\begin{equation}
  \frac{D^2\zeta^i}{d\tau^2} = \nu^2\,\zeta^i,
\qquad\textrm{namely,}\qquad
  \frac{d^2\zeta^i}{dt^2} = 0.
\end{equation}
%

\section{Geometric Point of View of the Newton-Hooke Limit}
\label{appdx:limit}

We have seen that the result of the limit (\ref{nu}) is not unique.  In
addition to Eq.(\ref{nu}), we need to specify the limits of the coordinates.
We should have also noticed that the topological structure of contraction may not
be the same as that of $\mathcal{B}_\Lambda$.  Here, we try to give a geometric
point of view of the contraction.

First, for a given positive $\Lambda = 3 R^{-2}$, $\mathcal{B}_\Lambda$ is
viewed as a hyperboloid of the 5-d Minkowski spacetime ${\cal M}^{1,4}$,
as shown in section~\ref{sect:BdS}.  When $\Lambda$, or $R$, runs over all the
positive values, a family of Beltrami-de~Sitter spacetimes are obtained.
Obviously, through each spacelike point in ${\cal M}^{1,4}$, there is
one and only one $BdS$ spacetime in the family.

Roughly speaking, to specify a particular limit evolving
$R\to\infty$, one must specify a region $D\subset {\cal M}^{1,4}$
consisting of spacelike points of ${\cal M}^{1,4}$ and a
congruence of curves in $D$ satisfying the following conditions. \\
(1)  Through each spacelike point in $D$ there is one and only one curve in the congruence.\\
(2)  At each spacelike point in $D$, the corresponding
curve in the congruence is not tangent to ${\cal B}_\La$
passing through the same point. Consequently, there is a direction
along each of these curves such that the parameter $R$ increases
strictly. That is, $R$ can be viewed as the
parameter for each curve in the congruence.\\
(3) On each curve, the parameter $R$ can take any sufficiently large positive
value.\\
(4) The congruence is smooth in the following sense.  The congruence of
curves can be described using four parameters $t$, $x^i$ ($i = 1$, 2, 3), say,
such that $t$, $x^i$ and $R$ is a set of local coordinates on $D$
when $R$ is sufficiently large.\\
Given such a region $D$ and a congruence $\mathscr{C}_D$ of curves
in $D$, which will be denoted by $(D,\mathscr{C}_D)$, these curves
can be viewed as the orbits of an associated limit process. Thus,
on a given curve $\gamma$ in the congruence $\mathscr{C}_D$,
points with different values of the parameter $R$ are viewed as
the ``same" point in the limit process. Each of these curves can
be identified with the limit point of the process. And the
parameters $t$, $x^i$ can be regarded as the coordinates of the
limit point.

So, different region or different congruence of curves give different limit
processes. In the following we will show some examples.

The first example is the NH limit obtained in section~\ref{NHlimit}.
The region $D$ is the set of spacelike points $\xi^A$ in ${\cal M}^{1,4}$
such that $\xi^4 > 0$. Obviously, $D$ is homeomorphic to $\mathbb{R}^5$.
The congruence $\mathscr{C}_{D,\nu}$ consists of curves
$\xi^A = \xi^A(R; t, x^1, x^2, x^3; \nu)$
with
\be
\ \{\xi^0, \xi^i,\xi^4\}
=(1 - \nu^2 t^2 + R^{-2}|\vect{x}|^2)^{-1/2}\{\nu t R, x^i, R \},
\ee
where $R$ is the parameter on the curve, $t$ and $\vect{x}$ are parameters
describing curves in $\mathscr{C}_{D,\nu}$, and $\nu$ is a parameter to
distinguish one congruence from another. Equivalently, the limit process
associated to the above congruence is to keep $t = \xi^0/(\nu\xi^4)$ and
$x^i = R\xi^i/\xi^4$ finite as $R$ tends to infinity. At the same time,
$c = \nu R$ also tends to infinity.


For the second example, let the region $D$ be the set of all spacelike points
in ${\cal M}^{1,4}$ and $\xi^A=\xi^A(R; T, \tilde{\vect{x}}; \nu)$ with
\be
\ \{\xi^0, \xi^i, \xi^4\}=(1 - \nu^2 T^2 + |\tilde{\vect{x}}|^2)^{-1/2}\{\nu T R, R\tilde{x}^i, R\}
\ee
then the congruence consisting of curves $\xi^A = \xi^A(R; T,
\tilde{\vect{x}}; \nu)$ corresponds to the limit process as shown
in section~\ref{unique}. In this limit process, $c = \nu R$ tends
to infinity while $T$ and the dimensionless spatial coordinates
$\tilde{\vect{x}}$ are kept finite. Obviously this also meets the
requirement of NH limit. The resulted spacetime is, once
again, a de Sitter spacetime. See, section~\ref{unique}.

Now let us consider the last example. The region $D$ will be the set of
spacelike points in ${\cal M}^{1,4}$ such that $\xi^3 > 0$.  In the limit
process, both $R$ and $c$ tend to infinity with $\nu = c/R$ kept fixed.
The region $D = D_{+1}\cup D_{-1}\cup D_{+2}\cup D_{-2}\cup D_{+4}\cup D_{-4}$
where $D_{\pm 4}$, for example, is the subset of $D$ on which $\xi^4$ is
positive/negative. Beltrami coordinates can be defined on each of these
subsets. For example, on $D_{\pm 4}$, the Beltrami coordinates can be
defined to be
\begin{equation}
  t = \frac{\xi^0}{\nu\xi^4}, \qquad
  x^a = \frac{\xi^a}{\xi^4}, \ (a = 1,\ 2), \qquad
  z = R\,\frac{\xi^3}{\xi^4}.
\end{equation}
Note that $\nu t$, $x^1$ and $x^2$ are dimensionless while $z$ has a dimension
of length.

We assume that all the coordinates are kept finite in the limit process. In
other words, we are considering a congruence of curves\footnote{
In fact, the parameters $t$, $x^a$ and $z$ could only describe part of the
curves in the congruence. These parameters corresponds to the Beltrami
coordinates on $D_{\pm 4}$. There exist some curves in the congruence that
must described by parameters corresponding to the coordinates on $D_{\pm 1}$,
$D_{\pm 2}$, respectively. %
} 
$\xi^A = \xi^A(R; t, x^1, x^2, z; \nu)$ on $D$ with
\be
\ \{\xi^0,\xi^1,\xi^2,\xi^3,\xi^4\}=\tilde{\sigma}^{-1/2}(t, x^1, x^2, z)\{\nu t R,x^1 R,
x^2 R,z, R \},
\ee
where
\begin{equation}
  \tilde{\sigma}(t, x^1, x^2, z) = 1 - \nu^2 t^2 + (x^1)^2 + (x^2)^2
    + \bigg(\frac{z}{R}\bigg)^2.
\end{equation}
The parameters $t$, $x^a$, $z$ and $R$ can be regarded to be coordinates
on $D_{\pm 4} \subset D$. The 5-d Minkowski metric is then
\begin{eqnarray}
  ds^2 & = & \frac{R^2}{\tilde{\sigma}}\,
  \Big(\nu^2 dt^2 - (dx^1)^2 - (dx^2)^2 - \frac{1}{R^2}\,dz^2\Big)
  + R^2\bigg(\sqrt{\tilde{\sigma}}\, d\frac{1}{\sqrt{\tilde{\sigma}}}\bigg)^2
\nonumber \\
  & & - \bigg(1 + \frac{z^2}{R^2\tilde{\sigma}}\bigg)\,dR^2
  + \frac{2z}{R\,\tilde{\sigma}}\,dz\,dR,
\end{eqnarray}
where %
$$
  \sqrt{\tilde{\sigma}}\,d\frac{1}{\sqrt{\tilde{\sigma}}}
  = \frac{1}{\tilde{\sigma}}\,\bigg(
  \nu^2 t\,dt - x^1\,d(x^1) - x^2\,d(x^2) - \frac{z}{R^2}\,dz
    + \frac{z^2}{R^3}\,dR\bigg).
$$

To obtain the limit, we first consider the induced metric on a hypersurface
$R = \const$, on which $dR = 0$. Then we take the limit of $c^{-2}ds^2$ as
$R\to\infty$. Obviously, its limit is
\begin{equation}
  d\tau^2 = \nu^{-2} g_{\alpha\beta}\,dx^\alpha\,dx^\beta,
\end{equation}
where $\alpha$ and $\beta$ take values from 0 to 2, with
$x^0 = \nu t$ and
\begin{equation}
  g_{\alpha\beta} = \frac{\eta_{\alpha\beta}}{\sigma_{BdS_3}(x)}
  + \frac{\eta_{\alpha\alpha'}\eta_{\beta\beta'}x^{\alpha'} x^{\beta'}}
    {\sigma^2_{BdS_3}(x)}, \qquad
  \sigma_{BdS_3}(x) = 1 - \eta_{\alpha\beta}\,x^\alpha x^\beta.
\end{equation}
Needless of speaking, we obtain a space-time having the structure
of $BdS_3\times\mathbb{R}$, where $BdS_3$ is a unit 3-d
Beltrami-de~Sitter spacetime, while $\mathbb{R}$ is labeled
by the coordinate $z$. %


\bigskip

\begin{acknowledgments}
The authors would like to thank Professors Y.~H.~Gao, Q.~K.~Lu,
 X.~C.~Song and Y.~Q.~Yu for valuable discussions.
Special thanks are given to Prof.~J.~Z.~Pan who made contribution
to part of this work. This work is partly supported by NSFC under
Grants Nos. 90103004, 10175070, 10375087, 10347148.
\end{acknowledgments}

\end{document}